\documentstyle[pre,aps,epsf,float,multicol]{revtex}
\begin{document}
\draft
\title{\bf Finite-Size Effects in the ${\bf \varphi}^{4}$ Field 
Theory Above the Upper Critical Dimension}
\author{X.S. Chen$^{1,2}$ and V. Dohm$^{1}$}
\address{$^1$ Institut f\"ur Theoretische Physik, Technische Hochschule Aachen,
D-52056 Aachen, Germany}
\address{$^2$ Institute of Particle Physics, Hua-Zhong Normal University, 
Wuhan 430070, China}
\date{September 8, 1997}
\maketitle
\begin{abstract}
We demonstrate that the standard $O(n)$ symmetric $\varphi^{4}$ field theory 
does not correctly  describe the leading finite-size effects 
near the critical point of spin systems on a $d$-dimensional lattice with 
$d > 4$. We show that these finite-size effects require a description 
in terms of a lattice Hamiltonian. For $n \rightarrow \infty$ and $n=1$ 
explicit results are given for the susceptibility and for the Binder 
cumulant. They imply that recent analyses of Monte-Carlo results for the 
five-dimensional Ising model are not conclusive.
\end{abstract}
\pacs{PACS numbers: 05.70.Jk, 64.60.Ak, 75.40.Mg}
\begin{multicols}{2}
The effect of a finite geometry on systems near phase transitions is of 
basic interest to statistical physics and elementary particle physics. 
In both areas the $\varphi^{4}$ Hamiltonian 
\begin{equation} \label{H}
H = \int_{V} d^{d} x \Bigl [{1\over 2} r_0 \varphi^{2} +
{1\over 2}(\bigtriangledown \varphi)^{2} + u_0 (\varphi^{2})^2  \Bigr ]
\end{equation}
for an $n$-component field $\varphi ({ \bf x})$ in a finite 
volume $V$ plays a fundamental role \cite{ZJ}. For simplicity we consider a 
$d$-dimensional cube, $V = L^{d} $, with periodic boundary conditions,
$\varphi ({ \bf x}) = L^{-d} \sum_{\bf  k} \varphi_{\bf  k}
e^{i {\bf  k} \cdot { \bf x}}\phantom{1}$.
The summation runs over discrete ${\bf  k}$ vectors with components
$k_j = 2 \pi m_j /L, m_j = 0,\pm 1, \pm 2, ... , j=1,2,... d,$ in the 
range $-\Lambda \leq k_j < \Lambda $ with a finite cutoff $\Lambda$.

It is generally believed that the leading finite-size effects near the 
critical point of $d$-dimensional systems can be described by $H$ 
both for $ d \leq d_u $ and for $ d > d_u $ where $d_u = 4 $ is the upper 
critical dimension. 
Since for $d > 4$ the bulk critical behavior is mean-field like, it is 
plausible that the leading finite-size effects for $d > 4$ appear to be 
describable in terms of a simplified Hamiltonian \cite{BZJ,RGJ}
\begin{equation} \label{H0}
H_{0} (\Phi) = L^{d} \Bigl [ {1\over 2}r_0 \Phi^{2}+ u_0 (\Phi^{2})^2 
\Bigr ]
\end{equation}
involving only the homogeneous fluctuations of the lowest $({\bf  k} = 0 ) $ 
mode $\varphi_0 = L^{d} \Phi$, $\Phi = L^{-d} \int_V d^d x \varphi ({ \bf x})$.
Based on the statistical weight $\exp [-H_0 (\Phi)]$, universal results have 
been predicted \cite{BZJ} for systems above $d_u$.
For the case $n=1$, the lowest-mode predictions have been compared with 
Monte-Carlo (MC) data for the five-dimensional Ising model [4-7]. Although 
disagreements were noted and doubts were raised in Ref.4, subsequent analyses 
[5-7] based on the Hamiltonian $H$ appeared to reconcile the MC data with the 
lowest-mode predictions.

In this Letter we shall demonstrate that the lowest-mode approach fails 
for the Hamiltonian $H$ in Eq.(\ref{H}) for $d > 4$ and that the leading 
finite-size effects of spin systems on a $d$-dimensional lattice with 
$ d > 4 $ are not correctly described by $H$. We show that this 
defect of $H$ is due to the $(\bigtriangledown \varphi )^{2}$ term. These 
unexpected findings shed new light on the role of lattice effects
for $d > d_u $ and imply that recent analyses of the MC data [4-7] in terms of
the continuum $\varphi^{4}$ theory are not conclusive.

We shall prove our claims first in the large-$n$ limit where a saddle point 
approach \cite{ZJ} can be employed. Our proof is not based on the 
renormalization group. We have extended the saddle point approach to the 
finite system to derive the order-parameter correlation function
\begin{equation}
\chi = {1\over n} \int_V d^d x <\varphi ({ \bf x}) \varphi (0)>
\end{equation}
with the statistical weight $\exp (-H)$.
In the limit $ n \rightarrow \infty $ at fixed $u_0 n$ we have found the exact
result
\begin{equation} \label{chi}
\chi^{-1}= r_0 + 4 u_0 n L^{-d} \sum_{\bf  k}( \chi^{-1} + {\bf  k}^2)^{-1}.
\end{equation}
We shall denote the bulk critical temperature by $T_c$. For $ T \geq T_c $, 
$\chi$ can be interpreted as the susceptibility (per 
component) of the finite system. In the bulk limit
the standard equation \cite{A} for the bulk susceptibility $\chi_{b}$ for 
$T \geq T_c$ is recovered from Eq.(\ref{chi}) as
\begin{equation} \label{chib}
\chi_{b}^{-1} = r_0 + 4 u_0 n \int_{\bf  k} \Bigl ( \chi_{b}^{-1} +
{\bf  k}^2 \Bigr )^{-1}
\end{equation}
where $\int_{\bf  k} $ stands for $(2 \pi)^{-d} \int d^d k $
with a finite cutoff $|k_j| \leq \Lambda$. 
It is convenient to rewrite Eq.(\ref{chi}) in terms of $r_0-r_{0c}=a_0 t$ 
where $r_{0c} = - 4 u_0 n \int_{\bf  k} {\bf  k}^{-2}$
is the bulk critical value of $r_0$ as determined from Eq.(\ref{chib}) 
( with $\chi_{b}^{-1} = 0 $), and $t=(T-T_c)/T_c$.
Furthermore it is important to separate the ${\bf  k} = 0$ part 
$4 u_0 n L^{-d} \chi$ from the sum in Eq.(\ref{chi}). After a simple 
rearrangement we obtain 
\begin{eqnarray}
\chi^{-1} &=&{\delta r_0 + \sqrt{(\delta r_0)^2+16u_0 n L^{-d}
(1+S) } \over 2 (1+S)}, \label{chi0}\\
\delta r_0 & = & a_0 t -\Delta , \label{dr0}\\ 
S &=& 4u_0 n L^{-d}\sum_{{\bf  k}\neq 0}
[{\bf  k}^{2}(\chi^{-1}+{\bf  k}^{2}) ]^{-1}, \label{S}\\
\Delta &=& 4 u_0 n \Bigl [\int_{\bf  k} {\bf  k}^{-2}
-L^{-d} \sum_{{\bf  k}\neq 0} {\bf  k}^{-2} \Bigr ].\label{Delta0}
\end{eqnarray}
These equations are the starting point of our analysis. They are exact in the
limit $n \rightarrow \infty $ at fixed $u_0 n $ and are valid, at finite cutoff
$\Lambda$, for $ d > 2$,
for arbitrary $L$ and for arbitrary $r_0$. 
They are written in a form that separates the ${\bf  k} = 0$ contribution 
$16u_0 n L^{-d}$ from the effect of the ${\bf  k} \neq 0$ modes. 
The latter is contained in $S$ and $\Delta$. 

In addition to the finite-size effect of the ${\bf k}=0$ mode, the ${\bf k} 
\neq 0$ modes cause two different finite-size effects:
(i) a finite renormalization of the coupling $u_0 n$ due to $S$ which for 
$d >4$ attains the finite bulk value $S_b = 4u_0 n \int_{\bf k} [{\bf k}^{2}
(\chi_b^{-1}+{\bf k}^{2})]^{-1}$, and
(ii) a shift of the temperature scale due to $\Delta$ which vanishes in the 
bulk limit. These two kinds of finite-size effects were also identified by 
Br\'{e}zin and Zinn-Justin \cite{BZJ} who argued that for $d > 4$ these effects
do not change the leading $L$ dependence obtained within the lowest-mode
approximation. These arguments do not depend on $n$ and, if correct, should 
remain valid also in the large-$n$ limit.

The finite-size effect (ii) comes from $\Delta$ which, for $d > 2$ 
and finite $\Lambda$, has the nontrivial large-$L$ behavior
\begin{equation} \label{Delta}
\Delta \sim 4 u_0 n \Lambda^{d-2}\Bigl [ a_1 (d) (\Lambda L)^{-2}+
a_2 (d) (\Lambda L)^{2-d} \Bigr ]
\end{equation}
apart from more rapidly vanishing terms. For the coefficients $a_{i} (d)> 0$
we have found
\begin{eqnarray}
a_1 (d) &=& {d \over 3(2\pi)^{d-2}} \int_{0}^{\infty} d x
x e^{-x} \Bigl [ \int_{-1}^{1} d y e^{-y^{2} x} \Bigr ]^{d-1},\label{a1}\\
a_2 (d) &=& {-1 \over 4 \pi^2} \int_{0}^{\infty} d y 
\Bigl [ (\sum_{m=-\infty}^{\infty} e^{-ym^2})^{d}-\Bigl ({\pi\over y}
\Bigr )^{d/2} -1 \Bigr ],\label{a2}
\end{eqnarray}
as confirmed in Fig.1 by numerical evaluation of Eq.(\ref{Delta0}) 
for $d=3,4,5$. Thus, for $d > 4$, $\Delta$ vanishes as $L^{-2}$, and not as 
$L^{2-d}$ [2,5-7] or as $L^{-d/2}$ \cite{RNB}. This implies that in 
Eq.(\ref{chi0}) the zero-mode term 
proportional to $L^{-d}$ does no longer constitute the dominant finite-size 
term.

Our claims are most convincingly examined at bulk $T_c $. 
Then Eq.(\ref{chi0}) is reduced to
\begin{equation} \label{chic}
\chi_{c}^{-1} ={-\Delta + \sqrt{\Delta^2+16u_0 n L^{-d}
(1+S_c) } \over 2 (1+S_c)}
\end{equation}
where $S_c$ is given by the r.h.s. of Eq.(\ref{S}) with $\chi^{-1}$ replaced 
by $\chi_{c}^{-1}$. We see that the large-$L$ behavior is 
significantly affected by the $\Delta^{2}$ term. For large $L$ and $d > 4$ 
we obtain from Eqs.(\ref{chic})and (\ref{Delta})
\begin{equation} \label{chica}
\chi_{c} \sim {L^{d} \Delta \over 4u_0 n } \sim a_1 (d) \Lambda^{d-4}
L^{d-2}.
\end{equation}
By contrast, the lowest-mode approximation with $\Delta = 0 $
and $S_c = 0 $ yields $\chi_{0c}=  (4 u_0 n)^{-1/2} L^{d/2}$.
This proves that the lowest-mode approach fails in the present case. We note
that the arguments in Ref.2 regarding the finite-size effect (ii) are 
not compelling since they are focused on the contributions of individual 
terms at lowest non-zero ${\bf k}$ rather than on an analysis of the 
{\it summed effect} of these contributions.
\begin{figure}
\narrowtext
\epsfxsize=\hsize\epsfbox{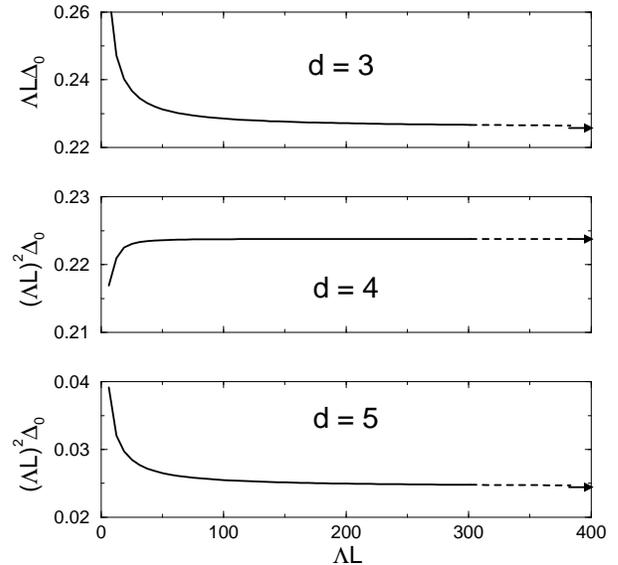}
\caption{$\Lambda L$-dependence of $\Delta_0 = \Delta /(4u_0 n 
\Lambda^{d-2})$ with $\Delta$ from Eq.(\ref{Delta0}) for $d=3,4,5$ 
(solid curves). The dashed lines represent Eq.(\ref{Delta}) 
with $a_1 (3) = 0.27706 $, $a_1 (4) = 0.08333$, $a_1 (5) = 0.02443$, and
$a_2 (3) = 0.22578$, $a_2 (4) = 0.14046$, $a_2 (5) = 0.10712$. The arrows
indicate the large-$\Lambda L$ limits.}
\end{figure}
Furthermore we see that the $L^{d-2}$ power law in Eq.(\ref{chica}) differs 
from the $L^{d/2}$ power law obtained from the exact solution of the 
$n$-vector model on a lattice for $n \rightarrow \infty$ \cite{B} and of the 
mean spherical model on a lattice \cite{SR}. This proves that the 
field-theoretic Hamiltonian $H$ in Eq.(\ref{H}) does not correctly describe 
the leading finite-size effects of spin models on a $d$-dimensional lattice 
with $d > 4$, at least in the large-$n$ limit.

In the following we show that this defect is due to the $(\bigtriangledown 
\varphi )^{2}$ or ${\bf k}^{2}\varphi_{\bf k}\varphi_{-\bf k}$ term of 
\begin{eqnarray} \label{HK}
H =&& L^{-d}\sum_{\bf k} {1\over 2} (r_0 + {\bf k}^{2}) \varphi_{\bf k}
\varphi_{-{\bf k} } \nonumber \\
&&+u_0 L^{-3d} \sum_{{\bf k}{\bf k}'{\bf k}''}
(\varphi_{\bf k}\varphi_{{\bf k}'})(\varphi_{{\bf k}''}
\varphi_{-{\bf k}-{\bf k}'-{\bf k}''})
\end{eqnarray}
with $\varphi_{\bf k} =\int_V d^{d} x e^{-i {\bf k}\cdot {\bf x}}\varphi 
({\bf x})$. Instead we consider a lattice Hamiltonian $\hat{H}(\varphi_{i})$ 
for $n$-component vectors $\varphi_{i}$ with 
components $\varphi_{i\alpha}$,
$-\infty \leq \varphi_{i\alpha} \leq \infty $, $\alpha = 1, ..., n$, 
on lattice points ${\bf x}_i$ of a simple cubic lattice with volume 
$V = L^{d}$ and with periodic boundary conditions. We assume 
\begin{equation} \label{HL}
\hat{H} (\varphi_i) =\tilde{a}^{d} \left \{ \sum_{i} \Bigl [ {\hat{r}_0 
\over 2}\varphi_{i}^{2} +
\hat{u}_0 (\varphi_{i}^{2})^{2} \Bigr ]+\sum_{i,j} {J_{ij}\over 2 
\tilde{a}^{2}} 
(\varphi_i-\varphi_j )^{2}\right \}
\end{equation}
where $J_{ij}$ is a pair interaction and $\tilde{a}$ is the lattice spacing. 
In terms of $\hat{\varphi}_{\bf k} = \hat{a}^{d}\sum_{j} 
e^{-i {\bf k}\cdot {\bf x}_j } \varphi_j$
the Hamiltonian $\hat{H}$ has the same form as Eq.(\ref{HK}) but with 
$r_0 + {\bf k}^{2}$ replaced by $\hat{r}_0 + 2 \delta J({\bf k})$ where
$\delta J({\bf k}) \equiv J(0)-J({\bf k})$ and
\begin{equation} \label{JK}
J({\bf k}) = (\tilde{a}/L)^{d}\sum_{i,j} J_{ij} e^{-i {\bf k}\cdot 
({\bf x}_i-{\bf x}_j)}.
\end{equation}
The ${\bf k}$ values are restricted by $-\pi/ \tilde{a}\leq k_j < \pi/
\tilde{a}$. In the large-$n$ limit at fixed $\hat{u}_0 n$ the susceptibility 
$\hat{\chi} = {1\over n}(\tilde{a}/L)^{d} \sum_{i,j} <\varphi_i \varphi_j > $ 
is determined by Eqs.(\ref{chi0})-(\ref{Delta0}) with ${\bf k}^{2}$ replaced 
by $2 \delta J({\bf k})$.
The large-$L$ behavior of the crucial quantity $\hat{\Delta}$ is for $d > 2$
\begin{eqnarray} 
\hat{\Delta} &=& 2\hat{u}_0 n \Bigl ( \int_{\bf k} [\delta J({\bf k})]^{-1}
-L^{-d} \sum_{{\bf k} \neq 0} [\delta J({\bf k})]^{-1} \Bigr )\label{DeltaL}\\
&\sim & 4\hat{u}_0 n J_0^{-1}a_2 (d) L^{2-d} \label{DeltaLa}
\end{eqnarray}
which for $d > 4$ differs from that of the continuum version $\Delta$, 
Eq.(\ref{Delta}), where $2\delta J({\bf k})$ was approximated by 
${\bf k}^{2}$. This approximation turns out to be the unjustified for $d > 4$.
Eq.(\ref{DeltaLa}) is valid for short-range interactions where
\begin{equation} \label{J0}
J_0 = {1\over d}(\tilde{a}/L)^{d}\sum_{i,j}(J_{ij}/\tilde{a}^{2})
({\bf x}_i-{\bf x}_j)^{2}
\end{equation}
is finite. As a consequence of Eq.(\ref{DeltaLa}), the leading $L$ dependence 
of $\hat{\chi}$ at $T_c$ is for $d > 4$
\begin{equation} \label{chicL}
\hat{\chi}_{c} \sim {1\over 2} (\hat{u}_0 n)^{-1/2}(1+S_c^{b})^{1/2} L^{d/2}.
\end{equation}
This agrees with the $L^{d/2}$ power law of the exact solution of the lattice 
models of Refs. 9 and 10 for $d>4$ and with the lowest-mode 
result $\hat{\chi}_{0c} ={1\over 2} (\hat{u}_0 n)^{-1/2} L^{d/2}$. We see that 
at $T_c$ for $d > 4$ the ${\bf k} \neq 0$ modes of $\hat{H}$ do not change 
the leading exponent $d/2$ of the lowest-mode approximation. Nevertheless they 
produce a finite change of the amplitude of $\hat{\chi}_{c}$ through
$S_c^{b} = \hat{u}_0 n \int _{\bf k} [\delta J({\bf k})]^{-2}$. Furthermore 
we conclude from Eqs.(\ref{chica}) and (\ref{chicL}) that for $d >4 $ the 
lattice Hamiltonian $\hat{H}$ yields significantly different finite-size 
effects compared to those of $H$.

An analysis of the temperature dependence of $\chi (t,L)$, Eq.(\ref{chi0}), 
and of $\hat{\chi} (t,L)$ shows that for $d > 4$ finite-size scaling in its
usual form is not valid, as expected \cite{B}, but we find that it remains 
valid in a generalized form {\it with two reference lengths}. The 
asymptotic scaling structure of $\hat{\chi}$ is 
\begin{equation}
\hat{\chi} (t,L) =
L^{\gamma/\nu}\hat{P}_{\chi}(t(L/\hat{\xi}_0)^{1/\nu}, (L/\hat{l}_0)^{4-d})
\end{equation} 
where $\hat{\xi}_0$ is the bulk correlation-length amplitude and $\hat{l}_0
 = [4\hat{u}_0 n J_0^{-2}(1+\hat{S}_c^{b})^{-1}]^{1/(d-4)}$ is a second 
reference length. The $d$-dependent scaling function reads
\begin{equation} \label{PL}
\hat{P}_{\chi} (x,y)=2J_0^{-1}\left \{ \delta (x,y)+\sqrt{[\delta (x,y)]^{2}
+4y}\right \}^{-1}
\end{equation}
where $\delta (x,y) = x-a_2 (d) y$. In the lowest-mode approximation the 
term $-a_2 (d) y$ is dropped which implies that the leading finite-size term
for $t > 0$ becomes incorrect. Thus, for $t > 0$, the lowest-mode approach 
fails for the lattice model (and also for the continuum model whose
scaling function $P_{\chi}$ turns out to be non-universal).

In the following we extend our analysis to the case $n = 1 $ which is of 
relevance to the interpretation of MC data of the five-dimensional Ising 
model [4-7]. We shall examine the susceptibility $\chi$ and the Binder 
cumulant $U$,
\begin{eqnarray}
\chi &=&\int_V d^{d} x <\varphi ({\bf x}) \varphi (0)> = L^{d} < \Phi^{2} >, 
\label{chi1} \\
U&=&1-{1\over 3} <\Phi^{4}>/<\Phi^{2}>^{2},\label{u1}
\end{eqnarray}
within the $\varphi^{4}$ model, Eq.(\ref{H}), including the effect of the 
${\bf k} \neq 0 $ modes in one-loop order. For $d > 4$ at finite cut-off, the 
perturbative finite-size field theory \cite{EDC,CDS} is applicable near 
$T_c$ without a renormalization-group treatment. 
The averages are defined as $<\Phi^{m}> = \int_{-\infty}^{\infty} d \Phi 
\Phi^{m} P (\Phi)$ where $P (\Phi) = Z^{-1} \int {\cal D} \sigma e^{-H}$ is 
the order-parameter distribution function with $\sigma ({\bf x}) = \varphi 
({\bf x}) -\Phi $ representing the inhomogeneous fluctuations \cite{CDS}. 
From Refs. 11 and 12 we derive the $L$ dependence 
at $T_c$ in one-loop order
\begin{eqnarray}
\chi_c &=& L^{d/2} u_0^{eff-1/2} \vartheta_2 (y_0^{eff}), \\ 
U_c    &=& 1-{1\over 3} \vartheta_4 (y_0^{eff})/\vartheta_2 (y_0^{eff})^{2},
\end{eqnarray}
where 
\begin{eqnarray}
y_0^{eff} &=& r_0^{eff}L^{d/2}u_0^{eff -1/2},\label{yef}\\
r_0^{eff} &=& r_{0c}+12u_0 S_1 (r_{0L})+144u_0^{2}M_{0}^{2}S_2 (r_{0L}), 
\label{ref} \\
u_0^{eff} &=& u_{0} - 36 u_0^{2} S_2 (r_{0L}),\\
r_{0L}    &=& r_{0c}+12u_0 M_{0}^{2},\\
M_0^{2} &=& (L^{d}u_0)^{-1/2}\vartheta_2 (r_{0c}L^{d/2}u_{0}^{-1/2}),\\
\vartheta_{m} (y) &=& {\int_{0}^{\infty} d s s^{m} \exp (-{1\over 2}y s^{2}
-s^{4} ) \over \int_{0}^{\infty} d s \exp (-{1\over 2}y s^{2}-s^{4} )}.
\end{eqnarray}
In this order the critical value $r_{0c} < 0$ is determined implicitly by the 
bulk limit ($r_0^{eff} = 0$) of Eq.(\ref{ref}),
\begin{equation}
r_{0c} = -12u_0 I_1(-2r_{0c})+36u_0 r_{0c}I_2 (-2r_{0c}),
\end{equation}
where $I_m (r)=\int_{\bf k} (r+{\bf k}^{2})^{-m}$.
The finite-size effect of the ${\bf k} \neq 0$ modes enters through 
\begin{equation}
S_{m} (r) = L^{-d} \sum_{{\bf k} \neq 0} (r+{\bf k}^{2})^{-m}.
\end{equation}
For large $L$ we have $r_{0L} = -2r_{0c}+{\cal O}(L^{-d})$ and
\begin{eqnarray} \label{refc}
r_{0}^{eff} = &-&12u_0 [I_1 (-2r_{0c})-S_1 (-2r_{0c}) ]\nonumber \\
&+& 36u_0 r_{0c} [I_2 (-2r_{0c})-S_2 (-2r_{0c})]+{\cal O}(L^{-d}).
\end{eqnarray}
Similar to $\Delta$ in Eqs.(\ref{Delta0}) and (\ref{Delta}), the parameter 
$r_0^{eff} < 0$ vanishes as $L^{-2}$ (rather than
as $L^{2-d}$) for $d > 4$, thus $y_0^{eff}< 0 $ {\it diverges} as 
$L^{(d-4)/2}$ (rather than {\it vanishes} as $L^{(4-d)/2}$) for $d > 4$.
Since $\vartheta_2 (y) \sim -y/4$ and $\vartheta_4 (y) \sim y^{2}/16$ for
large negative $y$ \cite{EDC} this implies that $\chi_{c}$ diverges as 
$L^{d-2}$ and $U_c$ attains the large-$L$ limit $2/3$. We conclude that 
the $L^{d/2}$ power law for $\chi_{0c}$ and the value $U_{0c}=1-{1\over 3}
\vartheta_4 (0)/\vartheta_2 (0)^{2}
=0.2705$ predicted \cite{BZJ} for $n=1$ within the lowest-mode approach 
are incorrect. From Refs. 2 and 12 we infer that analogous conclusions hold 
for general $n > 1$.

These unexpected results show that the widely accepted arguments in 
support of the asymptotic correctness of the lowest-mode approximation above 
the upper critical dimension in statics [1,2,5-7,13-19] and dynamics [1,20-23] 
are not valid and that recent interpretations [4-7] of the Monte-Carlo data 
of the five-dimensional Ising model in terms of predictions based on the 
Hamiltonian $H$, Eq.(\ref{H}), are not conclusive, in spite of the apparent 
agreement found in Refs. 5-7. 

Guided by our exact results in the large-$n$ limit, we propose a solution to 
this puzzle by replacing the field-theoretic $\varphi^{4}$ Hamiltonian $H$, 
Eq.(\ref{H}), by the lattice $\varphi^{4}$ Hamiltonian $\hat{H}$, 
Eq.(\ref{HL}), with $n=1$ for the comparison
with the five-dimensional Ising model. This involves a reexamination of 
$r_0^{eff}$, Eq.(\ref{refc}), with ${\bf k}^{2}$ replaced by 
$2[J(0)-J({\bf k})]$. We anticipate that the resulting value for the Binder 
cumulant $\hat{U}_c$ of the lattice model will be close to (or possibly 
identical with) that of the lowest-mode approach. This expectation is based 
on our result (for $n \rightarrow \infty$) that at $T_c$ the lowest-mode
approach yields the correct leading finite-size exponent of $\hat{\chi}_c$.
In addition, however, a detailed analysis of non-asymptotic (finite-$L$) 
correction terms is required which, for the Hamiltonian $\hat{H}$, are 
expected to be different from those employed in a recent analysis based on 
$H$ \cite{LB}. 

We summarize our findings for the continuum and lattice 
versions of the $\varphi^{4}$ model for $d > 4$ as follows: Lattice 
effects manifest themselves not only in changes of nonuniversal amplitudes 
but also in changes of the exponents of the leading finite-size terms as 
compared to the exponents of the continuum $\varphi^{4}$ model. 
The lowest-mode approach fails for the continuum $\varphi^{4}$ model, 
and also for the lattice model for $t > 0$. Therefore the values for the 
amplitude ratios derived previously \cite{BZJ} cannot be justified on the 
basis of the 
$\varphi^{4}$ continuum theory. For the lattice Hamiltonian, however, the 
lowest-mode approach is qualitatively justified at $T_c$ for $n \rightarrow 
\infty $, at least for $\hat{\chi}_c$. We conjecture that this is the reason 
for a fortuitous (approximate) agreement found between MC (lattice) data [4-7]
 and the lowest-mode predictions \cite{BZJ}. Further work is necessary in 
terms of the $\varphi^{4}$ lattice model, Eq.(\ref{HL}), to fully establish
our conjecture. 

We also anticipate lattice and cutoff effects on leading finite-size terms 
at $d = d_u $. This is relevant to future studies of tricritical 
phenomena at $d=3$, e.g., in $^3$He-$^4$He mixtures \cite{ML} and to MC 
simulations for lattice models of elementary particle physics at $d=4$.

Support by Sonderforschungsbereich 341 der Deutschen 
Forschungsgemeinschaft is acknowledged.

\end{multicols}
\end{document}